\documentclass{ws-procs9x6}

\begin{document}

\title{The Search for AGN in Distant Galaxy Clusters
}

\author{R. DOWSETT, O. JOHNSON, P.N. BEST}

\address{Institute for Astronomy \\
Royal Observatory, Blackford Hill \\
Edinburgh, EH9 3HJ, UK \\
E-mail: red@roe.ac.uk}

\author{and O. ALMAINI}

\address{The School of Physics and Astronomy \\
University of Nottingham \\
Nottingham, NG7 2RD, UK }

\maketitle

\abstracts{We are undertaking the first systematic study of the
prevalence of AGN activity in a large sample of high 
redshift galaxy clusters. 
Local clusters contain mainly red elliptical galaxies, and have little or
no luminous AGN activity. However, recent studies of
some moderate to high redshift clusters have revealed significant numbers
of luminous AGN within the cluster. 
This effect may parallel the Butcher-Oemler effect --  the increase in
the fraction of blue
galaxies in distant clusters compared to local clusters.
Our
aim is to verify and quantify recent evidence that AGN 
activity in dense environments increases with redshift,
and to evaluate the significance of this effect. 
As cluster AGN are far less prevalent than field sources, 
a large sample of over 120 cluster fields 
at $z > 0.1$ has been selected from the Chandra archives,
and is being analysed for excess point sources. The 
size of the excess, the radial distribution and flux 
of the sources 
and the dependence of these on cluster redshift and luminosity 
will reveal important information about the triggering and
fueling of AGN. }

\section{Status of Cluster AGN Studies}

X-ray images are one of the most efficient ways to detect moderate to high
redshift AGN. They are also highly effective for detection of galaxy
clusters, through the diffuse emission from the intra-cluster medium. In
one image it is possible to identify both the properties of the cluster and
the probable number of X-ray sources associated with it by comparison to a
blank field.

We have analysed two such clusters, MS1054-0321\cite{1} (see O. Johnson et. al,
this proceedings) and MS1512+36. Both clusters show a statistical excess of
sources. MS1054-0321, a rich cluster at $z=0.83$, has an excess of
luminous sources in the outer regions (1 - 2 Mpc from the cluster
centre). In contrast MS1512+36 ($z=0.37$) is a far poorer cluster and
its excess~sources are found in the central 1Mpc, and are an order of magnitude
less~luminous. 

In addition to these two clusters, excess AGN have been found in the
regions of five other clusters  and two probable protoclusters (see
discussion in Johnson et al.\cite{1} 2003), spanning $0.15 < z < 2.16$. There is
only one documented case of a high redshift cluster with no excess AGN\cite{2} - however, many null results may go unpublished.  In many cases
spectroscopic and photometric observations have confirmed that the excess
AGN do lie within the cluster\cite{3}.  From these seven clusters there is some
evidence for evolution of the typical luminosity of the cluster AGN with
redshift. The lower redshift clusters $(z < 0.3)$ have excesses dominated
by low luminosity sources. In the higher redshift clusters a population of
more luminous sources emerges, but the small sample size makes any
conclusions highly speculative.

\vspace*{-0.3cm}

\section{Our Project: A Systematic Survey of Galaxy Clusters}

What is clearly required now is a systematic study of a large, well defined
sample of galaxy clusters in order to quantify the prevalence and form of
AGN activity in dense environments at intermediate and high redshifts. We are
undertaking a major project to study over 120 clusters at $z >
0.1$ from the Chandra X-ray Observatory archive. We have 
developed an automated pipeline to reduce the data, detect sources and
analyse the results. 

Our project will focus on the statistical excess of point sources in
cluster fields compared to blank fields. Extensive calibration is being
undertaken to determine the degree of field to field variation in
non-cluster fields, and to identify any small systematic effects that could
dominate over many fields. 
We will investigate the trends in the excess of point sources
with cluster redshift, luminosity and morphology. In addition our automated
pipeline analyses AGN luminosity and radial position within the cluster (as
described for the two examples above) for all the clusters in the sample.

With this major survey, we will determine how the properties of cluster AGN
depend on redshift and environment. This will identify some of the
mechanisms that trigger or suppress AGN activity.


\end{document}